# Unlocking the full potential of jumping condensation on microstructured surfaces

*Mariia S. Kiseleva[1,4*], Tytti Kärki[1,4], Mika Latikka[1,4], Parham Koochak[1,4], Sakari Lepikko[1,4], Maja Vuckovac[1,4], Tomi Koskinen[2], Ramesh Raju[2], Ville P. Jokinen[3], Jiazheng Liu[5], Nenad Miljkovic[5,6,7,8,9,10], Jaakko V.I. Timonen[1,4], Ilkka Tittonen[2], and Robin H. A. Ras[1,4*]*

1. Department of Applied Physics, Aalto University, P.O. Box 15100, 02150 Espoo, Finland
2. Department of Electronics and Nanoengineering, Aalto University, P.O. Box 13500, 02150 Espoo, Finland
3. Department of Chemistry and Materials Science, Aalto University, P.O. Box 16100, 02150 Espoo, Finland
4. Centre of Excellence in Life-Inspired Hybrid Materials (LIBER), Aalto University, P.O. Box 15100, 02150 Espoo, Finland
5. Department of Mechanical Science and Engineering, The Grainger College of Engineering, University of Illinois Urbana-Champaign, Urbana, IL 61801 USA
6. Materials Research Laboratory, University of Illinois Urbana-Champaign, Urbana, IL 61801, USA
7. Department of Electrical and Computer Engineering, University of Illinois Urbana–Champaign, Urbana, IL 61801, USA
8. International Institute for Carbon Neutral Energy Research (WPI-I2CNER), Kyushu University, 744 Motooka, Nishi-ku, Fukuoka, 819-0395, Japan
9. Air Conditioning and Refrigeration Center, Department of Mechanical Science and Engineering, University of Illinois Urbana-Champaign, Urbana, IL, USA
10. Energy and Environment (iSEE), Institute for Sustainability, University of Illinois, Urbana, IL, USA

*Corresponding authors: mariia.kiseleva@aalto.fi, robin.ras@aalto.fi

**Funding**: Tandem Industry Academia from the Finnish Research Impact Foundation (Vaikuttavuussäätiö) project No. 239, Business Finland grant No. 5496/31/2024 (R.H.A.R, M.S.K.) and 5714321 (M.V.), Academy of Finland project No. 346109, No. 342169 (R.H.A.R.) and No. 341459 (V.J.), Academy of Finland Research Fellowship No. 361541 (M.V.), International Institute for Carbon Neutral Energy Research (WPI-I2CNER) (N.M.).

**Abstract**

Water condensation on superhydrophobic surfaces can generate spontaneous droplet jumping, enabling rapid condensate removal and improved thermal and mass transfer. Although this effect has been extensively demonstrated on densely packed nanostructures, the capability of microscale textures to support jumping condensation remains poorly understood. Here, we show that engineered microscale conical arrays can achieve efficient microdroplet jumping and reveal a previously unreported spacing-dependent critical transition between jumping and non-jumping regimes. In the jumping regime, by varying only the cone pitch, we identify a geometric threshold below which sub-10 µm droplets are rapidly removed, and above which jumping is suppressed, resulting in slower dynamics and larger departing droplets. In-situ optical and environmental scanning electron microscopies reveal the mechanistic origin of this transition: dense arrays favour full Cassie droplets, which depart cleanly, while wider spacing favours partial Cassie droplets that retain a localized wet region initiating new nucleation. From these results, we construct a geometry–wetting design map linking microstructure spacing, droplet morphology, and nucleation density. These findings establish design principles for scalable, mechanically robust microstructured surfaces capable of high-performance condensation management for anti-fogging, water harvesting, and heat-transfer applications.

## 1. Introduction

Coalescence-induced droplet jumping on superhydrophobic surfaces provides an efficient mechanism for condensate removal. When neighboring droplets merge, excess surface energy propels the droplet away,[1] initiating repeated cycles of nucleation, droplet growth, coalescence, and departure. This process minimizes surface coverage, making it attractive for anti-dew,[2,3] anti-icing,[4] anti-biofouling,[5] and anti-particulate fouling[6,7] applications. Continuous renewal of nucleation sites also enhances heat transfer[8–10] and accelerates water collection.[11] Despite these advantages, practical implementation is limited. Droplet dynamics are challenging to control, and eventual flooding of surface cavities leads to pinned Wenzel droplets and diminished jumping efficiency.[9,12,13]

Surface geometry strongly influences droplet nucleation, growth, morphology, and removal. Jumping condensation has been demonstrated on knife-like nanostructures,[9,13] micropillars,[14,15] and other microtextures.[16,17] Dense nanocone arrays, inspired by cicada wings, can efficiently shed sub-10 μm droplets,[2,3] but their fabrication is complex and resource-intensive.[18] Stochastic nanostructures are easier to produce[9,19,20] but offer limited control over feature geometry. In contrast, microscale texturing provides a scalable route to robust superhydrophobic surfaces, compatible with maskless etching[21] and high-resolution 3D printing.[22] Understanding condensation on microscale architectures is therefore essential for translating coalescence-induced jumping to real-world applications.

Droplet growth morphology further governs condensation performance. On microstructured surfaces, droplets can adopt suspended (full Cassie) or partially wetted (partial Cassie) states.[15] Both can jump, but they differ in growth rates, nucleation behavior, and heat-transfer characteristics. Partial Cassie droplets grow faster and initiate new nucleation,[23] influencing subsequent jumping dynamics and droplet size distribution.

Here, we combine well-defined microconical arrays with in-situ optical, confocal, and environmental electron microscopies to track microdroplet formation and evolution. By systematically varying cone spacing, we reveal how microstructure density controls droplet morphology, coalescence-induced jumping, and overall condensation performance. For the first time, we identify a sharp, spacing-dependent transition between jumping and non-jumping regimes on microscale conical arrays. In the jumping regime, sub-10 μm droplets are efficiently removed below the critical spacing and suppressed above it. We translate these findings into a quantitative geometry–wetting design map that links microstructure spacing, droplet morphology, and nucleation density, providing practical guidelines for engineering mechanically robust surfaces for anti-fogging, water harvesting, and heat-transfer enhancement.

## 2. Results and discussion

### 2.1. Cones geometry and wetting characterization

To investigate the effect of surface geometry on jumping dynamics and droplet morphology, we chose sharp microcones of uniform size varying only spacing $p$ (peak-to-peak) between cones (**Figure 1a** and **Table S1**). We prepared a series of samples with regular, hexagonally packed cones ($p$ = 1 – 4 µm), as well as irregular, densely packed array of sharp cones (so-called 'black silicon', $p$ < 0 .78 µm). For all samples we calculated solid fraction $\varphi$ and surface roughness $R_f$ (**Table S1**). All microstructured samples (**Supplementary note 1** on sample preparation) were coated with perfluorooctyltrichlorosilane (PFOTS) by chemical vapor deposition (CVD) method.[24] The coated planar surface exhibited advancing/receding contact angles of $\theta_a / \theta_r$ =117.9 ± 0.56 / 110 ± 0.3° with contact angle hysteresis $CAH$ = 8 ± 1.6°. By combining microstructuring and chemical modification, all samples became superhydrophobic with contact angle higher than 160° (**Figure S1a,** and **Supplementary note 2** on Contact angle measurement). Additionally, we used a scanning droplet adhesion microscope – SDAM (**Supplementary note 3**)[25] to measure the adhesion of droplets with a diameter of 1.4 mm (**Figure S1b**). Adhesion force scaled with the solid fraction, with smaller adhesion corresponding to sparser cones, consistent with previously reported friction measurements.[26] Macroscopic drops used for wetting characterization primarily interact with the tops of cones. Consequently, denser regular cone arrays lead to higher contact angle hysteresis and stronger adhesion forces, while 'black silicon' with hierarchical nanostructures on microcones exhibit lower hysteresis and adhesion due to their complex multiscale surface topography.

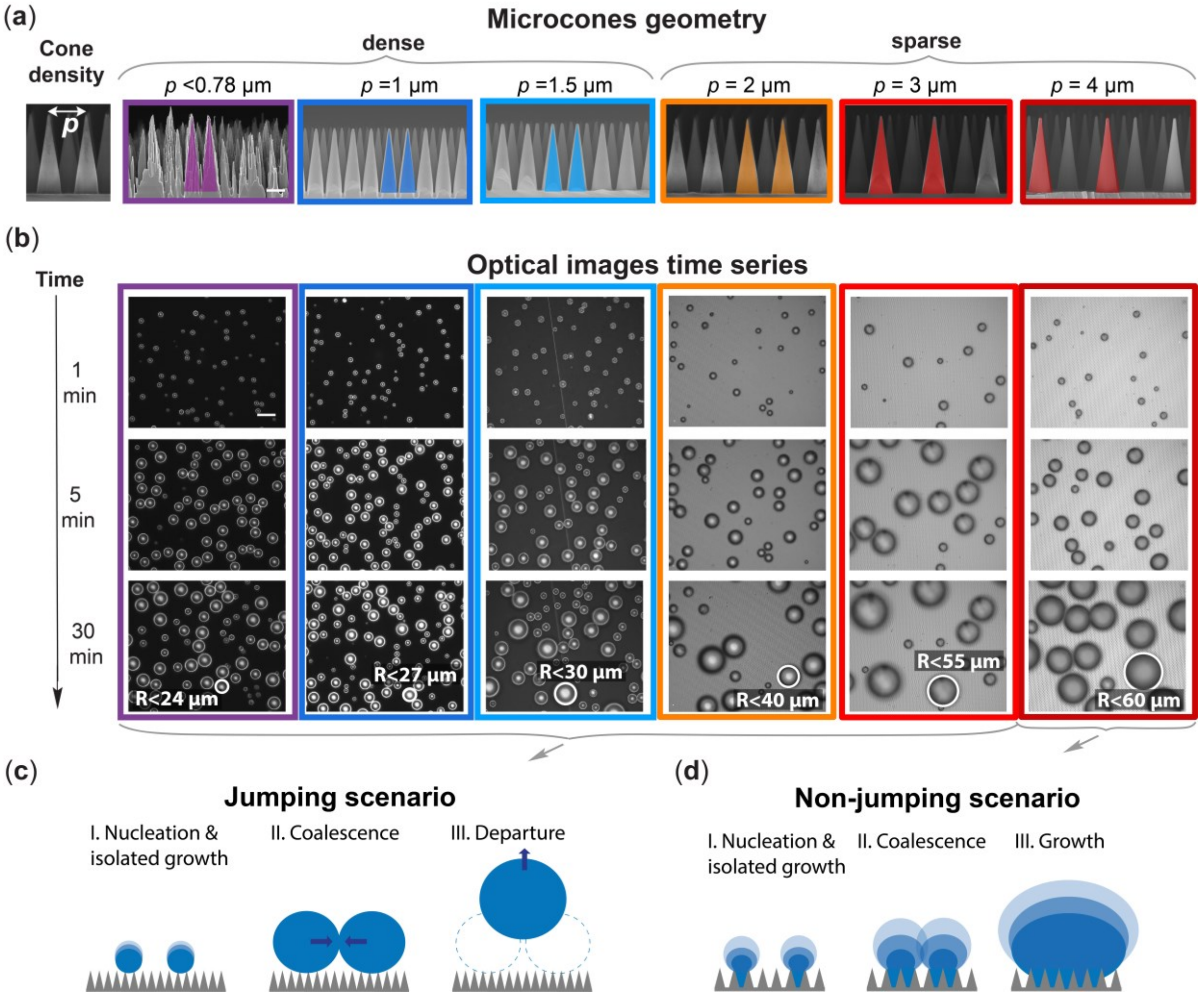

Figure 1. **Evolution of condensation on microcones: visualization and main scenarios *(a)*** SEM micrographs of dense irregular cones with spacing $p$ < 0.78 µm (violet) and hexagonally packed cones with spacing $p$ = 1, 1.5, 2, 3, and 4 µm (blue, light blue, orange, red, dark red, respectively). Samples with $p$ < 2 µm are referred to as dense, and those with $p \geq 2$ µm as sparse. Scale bar is 1 µm. (***b***) Time series of breath figures at 1, 5, and 30 minutes of the condensation process on an array of microcones. Maximum droplet radii after 30 min of experiment are indicated with white circles. Scale bar is 50 µm, which applies to all images in the panel. Schematics of the dominant droplet coalescence scenario for cones with (***c***) $p$ = 0.78, 1, 1.5, 2, 3 µm, which show droplet jumping, and for cones with (***d***) $p$ = 4 µm, which show no jumping.

## 2.2. Condensation performance and jumping dynamics

Condensation patterns on prepared microconical surfaces were visualized using an inverted portable microscope in controlled ambient temperature and humidity (**Figure S3** and **Supplementary Note 4**). In the time series of optical images, we observed the evolution of breath figures on surfaces with variable density of the cones (**Figure 1b**; **Supporting video 1**). Time series were analyzed with a custom-made image processing algorithm (**Supplementary note 5**) tracking each condensing droplet and defining its radius, position, circularity, etc. The postprocessing algorithm enabled identification of merging and jumping events during the experiment.

Samples with densely packed cones having spacing $p < 2$ µm ('black silicon', hexagonally packed cones with 1 and 1.5 µm spacing) will be referred to as dense cones, while samples with $p \geq 2$ µm will be referred to as sparse cones (2, 3, and 4 µm spacings). Dense cones appeared black due to high absorption of incident light. Conversely, sparse cones appear grey due to light reflection from flat areas between the cones, with individual cones being clearly visible on micrographs.

The typical condensation cycle includes three steps: (1) nucleation of multiple randomly distributed droplets, (2) growth of isolated droplets without coalescence, (3) coalescence of neighbouring droplets, which may result in either merging or jumping. In the jumping scenario (**Figure 1c**), after coalescence, two or more droplets gain sufficient kinetic energy to depart from the surface. Upon departure, the cycle repeats continuously. In the merging scenario (**Figure 1d**), the resulting droplet continues to grow until the next coalescence.

All samples except sample with 4 µm spacing exhibited jumping condensation scenario. In the early stage of condensation process, small droplets may merge due to strong pinning at the local defects, but resulting droplets will be eventually removed by the jumping mechanism. In contrast, on cones with 4 µm spacing condensing droplets were unable to jump, they merge and remain on the surface instead behaving similarly to condensing droplets on flat untextured surface.

**Figure 2a** shows that the nucleation density, $N_s$ (defined here as the droplet number per projected surface area), decreases with increasing spacing. Although nucleation occurs at the nanoscale and optically observed droplets form through coalesced nanodroplets, we use the term 'nucleation density' ($N_s$) for simplicity, instead of the more precise droplet number density (*DND*). Maximum (peak) values of $N_s$ for all samples from dense to sparse were 1500 ± 1290, 754 ± 352, 657.2 ± 337, 297.8 ± 285.2, 141.1 ± 95.1, 129.5 ± 24.8 $mm^{-2}$, respectively. For comparison, the maximum $N_s$ for planar silicon functionalized with PFOTS using the same deposition protocol under similar condensation conditions was 350 ± 40 $mm^{-2}$.[24] Variation in nucleation density with feature density is not sufficiently discussed in the literature due to large variation in surface design, mostly irregular nanostructured,[13,23] and inconsistent experimental conditions.[27,28] We suggest that in microscale roughness cases, higher nucleation density on dense cones can be explained by the larger total surface area compared to sparse cones, and the increased feature density, which enhances the chances of nanoscopic droplets merging to form a stable droplet. In contrast, on sparse cones having a smaller total area, nanoscopic droplets are less likely to encounter other nanodroplets.

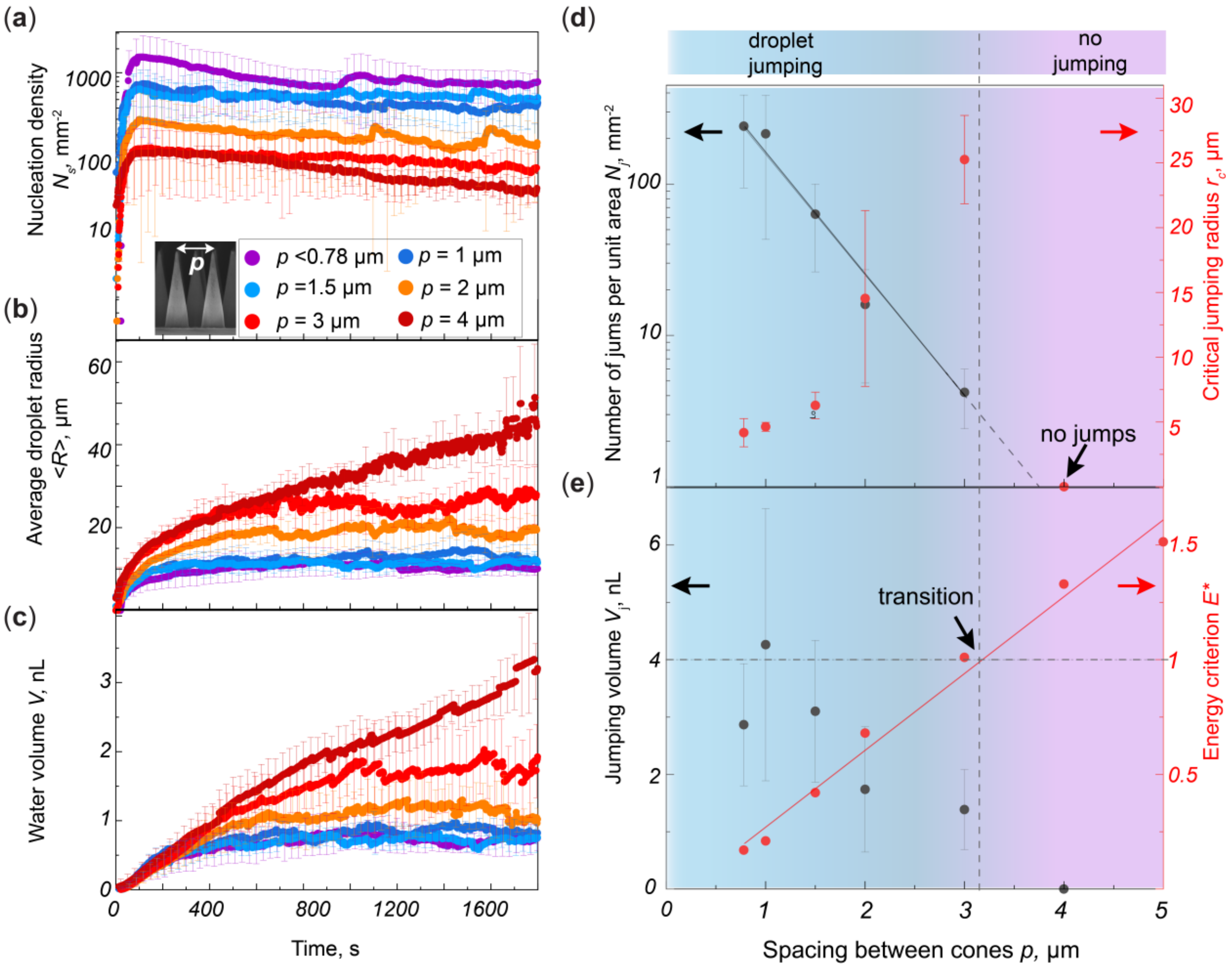

Figure 2. **Droplet jumping efficiency on microcones: from active to inhibited jumping.** Kinetics of (***a***) droplet number density in logarithmic scale, (***b***) average droplet radius <*R*>, (***c***) volume of remaining water on the sample. Insert in (a) depicts spacing *p* between cones; Same legend for panels (a), (b), (c). Error bars correspond to the standard deviation of at least 4 independent measurements ($n$ = 4…5) on different samples at similar conditions. (***d***) Number of droplet jumps per unit area in logarithmic scale (left side) and critical jumping radius $r_c$ (right side); (***e***) jumping volume $V_j$ (left side) and dimensionless energy criterion $E^*$ (right side) as a function of spacing *p*. To quantify the total number of events, a total experimental time of 30 minutes was used.

Active droplet removal on samples with jumping condensation keeps surface coverage $\Phi$ (proportion of apparent area covered by droplets to projected imaging area) below 35% with rare spikes up to *ca.* 40% for sparse cones (**Figure S4a**). The maximum droplet radius growth is limited by jumping condensation (**Figure S4b**). The average droplet radius <*R*> of condensed droplets is significantly smaller for dense cones compared to sparse ones, i.e. *ca.* 10 µm vs. 22–46.2 µm (**Figure 2b**), respectively. Since the volume of remaining water on the sample *V* is proportional to the cube of the droplet radius (the droplets are mostly spherical, as will be described later), larger droplets contain greater volumes despite their lower distribution density. Consequently, *V* is significantly larger on sparse cones than on dense ones (**Figure 2c**).

The kinetic curves of <*R*> and *V* exhibit similar shapes, consisting of linearly increasing segments associated with droplet nucleation, growth, and coalescence, followed by a plateau caused by droplet removal through jumping (small spikes on the plateau curves are attributed to droplet removal events). Jumping condensation limits both the droplet size and total condensed volume. In contrast, for samples with 4 μm cones, where no jumping occurs, <*R*> and *V* increase linearly.

We extracted the frequency of droplet removal and the size of removed droplets by analyzing droplet jumping dynamics. We found that number of jumps $N_j$ decreased exponentially (linearly in log scale) with increased spacing between cones, reaching zero for cones with a spacing of 4 μm. The droplet jumping rate *J*, defined as the proportion of jumping events relative to all coalescence events, increases steeply for dense cones (**Figure S5**), indicating that droplets are removed at the very early stage of condensation. On the contrary, on sparse cones, droplets undergo a merging stage, delaying jumping. The critical jumping radius $r_c$ (radius of droplets at which the jumping rate reached 50%) (**Figure 2d** and **Figure S6)**, increasing from *ca.* 5 μm for dense cones to 15 and 25 μm for sparse cones, while there is no jumping for the sample with 4 μm spacing. The jumping volume $V_j$ (**Figure 2e**), which can be considered as an integral parameter considering the number of jumps as well as the radii of jumping droplets, goes in line with the number of jumps trend: $V_j$ decreases linearly while spacing increases. Notably, for both dense cones and the 2 μm spacing sample, $V_j$ exceeds the volume of remaining condensed water on the sample (**Figure S7**), whereas for the 3 μm and 4 μm spacing samples, the volume of remaining condensed water is greater.

Trends for $N_j$ and $V_j$ align very well with predictions of energy criterion[29] (**Figure 2e** right side, **Table S1**) $E^* = cos\theta_a^{CB}/cos\theta_a^W = -1/(R_f cos\theta_a)$, where $cos\theta_a$ – cosine of the intrinsic advancing contact angle, and $R_f$ is the surface roughness. For $E^*$ > 1, the Wenzel state is favorable, while $E^*$ < 1 Cassie state prevails. A transition point occurs at $E^*$ = 1, corresponding to a 3 μm spacing, where jumping still occurs, unlike for the next sample with 4 μm spacing. Predicted by simple theoretical parameter $E^*$ transition in droplet morphology can shed the light into changes of condensation performance with increasing the spacing and explain failure of jumping for sample with 4 μm spacing.

### 2.3. Studies of microdroplet morphology

To explain the difference in jumping performance between dense and sparse cones, we analyzed microdroplet growth using environmental scanning electron microscopy (ESEM) (**Supplementary Note 6**). On dense cones, especially 'black silicon', very small droplets reached contact angles above ~170°, enabling early-stage removal (**Figure 3a**). In contrast, droplets on sparse cones required larger radii to exceed the ~160° threshold for jumping.[14] ESEM further revealed two distinct morphologies: [15] Cassie droplets (**Figure 3a**, insert, **Figure S8, a-c**), and partial Cassie droplets (**Figure 3a**, insert, **Figure S8d**). Partial Cassie droplets dominated on sparse cones, consistent with their reduced jumping efficiency.

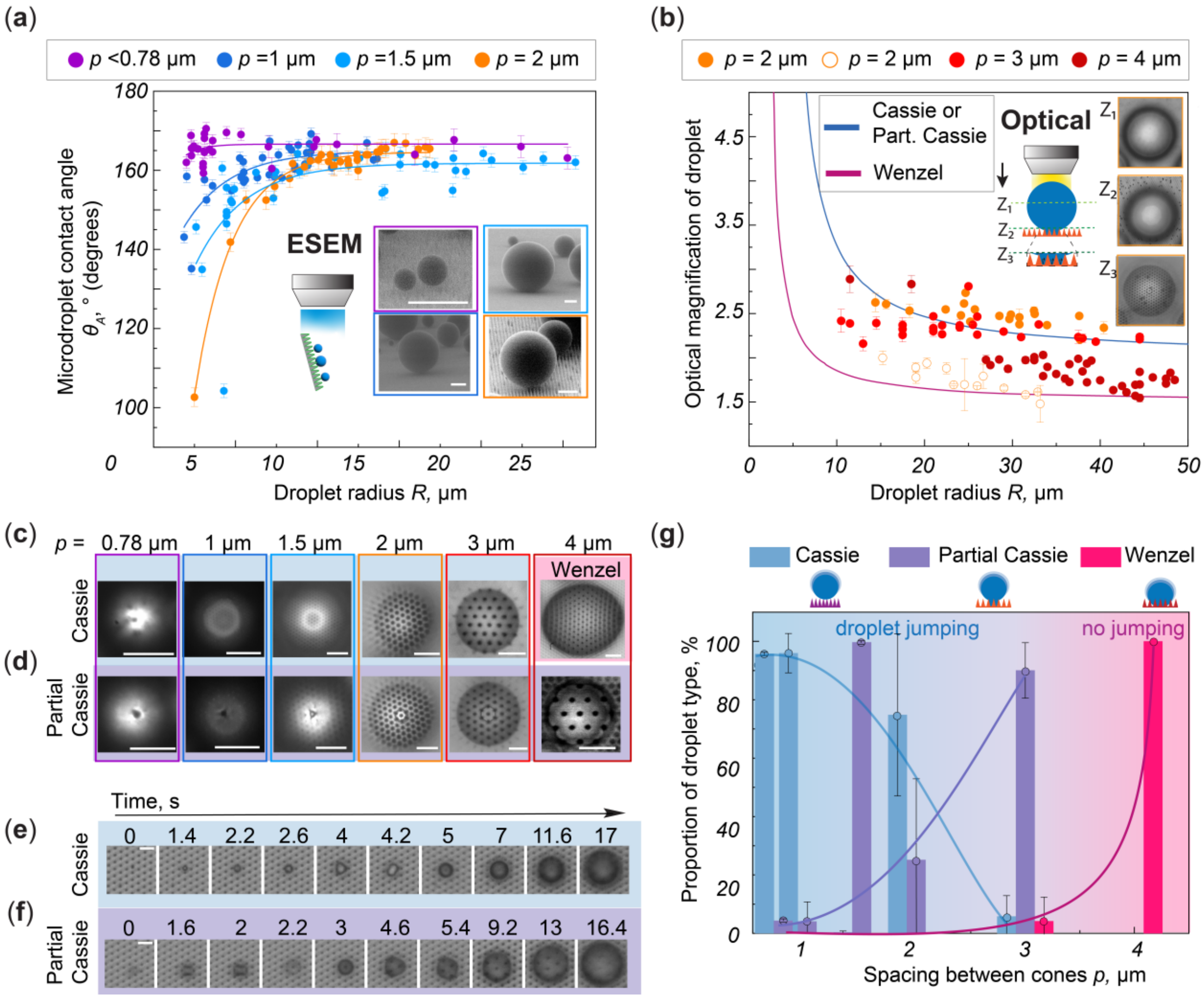

Figure 3. **Transitions in microdroplet morphologies: from jumping Cassie to non-jumping Wenzel droplets towards intermediate partial Cassie state.** (***a***) Microdroplet visualization using environmental SEM: contact angle as function of droplet radius, standard deviation was defined from 10 measurements of the same droplet. The inserts are representative images of microdroplets on surfaces with variable spacing between the cones. Scale bar is 10 µm. (***b***) Magnification effect of condensing droplets on sparse cones ($p$ = 2, 3, 4 µm; hollow orange circles correspond to failed sample with $p$ = 2 µm exhibiting no jumping). Samples with $p$ = 2, 3 µm follow Cassie or partial Cassie magnification model (blue solid line), whereas samples with $p$ = 4 µm and failed sample with $p$ = 2 µm follow Wenzel model (magenta solid line). Insets show schematic illustration of the optical setup used to image droplets at different focal planes ($Z_1$, $Z_2$, $Z_3$) and perform magnification analysis. Optical microscopy visualization of microdroplets in (***c***) Cassie, (***d***) partial Cassie having fully wetted area surrounded by air layer, and Wenzel states on samples with variable spacing. Early stages of droplet growth on sparse cones with (***e***) Cassie and (***f***) partial Cassie morphologies. Scale bar is 4 µm. (***g***) Proportion of each droplet type after 30 min of experiment defined by changing focus, solid lines (blue, violet and pink) are guide for the eye presenting global trend across range of samples. Error bar is standard deviation calculated from at least 4 measurements ($n$ = 4…20).

Along with ESEM we used non-invasive optical imaging from the top, varying the focal plane of the microscope,[30,31] we were able to visualize the contact area of the droplet and the surface (**Figure 3b,** insertion). The advantage of 'sparse' arrangement is that light can be reflected from the flat area between bases of adjacent cones, whereas densely packed cones prevent such reflection. Optical microscopy of droplet morphology was validated by magnifying factor analysis considering optical models of Cassie and Wenzel droplets (**Figure S9-11**), where condensing droplets act as magnifying lenses (**Figure 3b**, **Supplementary Discussion 1**), and further confirmed using confocal laser scanning microscopy (**Supplementary Note 7**). The obtained optical imaging data corroborated the ESEM measurements: we observed Cassie (**Figure 3c**), partial Cassie (**Figure 3d**), and Wenzel (**Figure 3c**) droplets.

When optically scanning a Cassie droplet (**Figure 3c**) by varying focal plane, the interface remains continuous as the droplet rests on cone tops with air beneath. In contrast, scans of a partial Cassie droplet (**Figure 3d**) show clear intensity variations revealing a circular or triangular wetted spot with defined boundaries, marking the transition from a water–air to a water–solid interface, observed at different focal planes for different droplet sizes. For dense cones stable configuration of wetted spot (**Figure 3d**) is a triangle between adjacent cones, while for sparse ones (2 and 3 μm spacing) it is hexagon with one cone inside, and almost a circle of 16 μm in diameter for 4 μm spacing sample. Interestingly, on non-jumping sample with 4 μm spacing condensing droplets are initially in partial Cassie that collapse into Wenzel state after the first merging event.

Using varying focal-plane scanning, we constructed a diagram illustrating the interaction between two droplets with distinct morphologies in the Cassie, partial Cassie, and Wenzel states (**Figure S12** and **Supplementary Discussion 2**). Partial Cassie droplets with one wetted spot were most commonly observed during scanning, although we also observed droplets with two wetted spots or enlarged wetted areas formed after merging near defects. While the evolution of these wetted regions and their transition to the Wenzel state is beyond the present scope, it may offer insight into jumping failure and the flooding phenomenon limiting jumping condensation applications.[12,32]

Assuming that the wetted spot represents the contact area between the surface and droplet (**Figure S13a, c**), we calculated the contact angle of the microdroplet and compared it with ESEM data (**Figure S13d**). Both methods exhibit similar trends, though optical estimates yield slightly higher angles. Notably, all samples including the non-jumping with 4 µm cone spacing sample bypass the threshold 160° contact angle indicating the ability to jump after droplet coalescence. For the 4 µm cone sample, a partial Cassie droplet would need to reach a radius of *ca.* 24 µm to jump. However, after merging, the droplet transitions to the Wenzel state. Even if a partial Cassie droplet reached the critical radius without coalescence, neighboring droplets already in the Wenzel state would prevent jumping upon merging. We defined the critical jumping radius from optical data $r_c^{opt}$ as the minimum radius where the contact angle exceeds 160°, and compared it with the experimental $r_c$ and the coalescence radius $R_e = 1/(4\sqrt{N_s})$ (**Figure S13d**). The lower $r_c^{opt}$ values relative to $r_c$ suggest potential for optimizing the jumping radius by increasing the nucleation density $N_s$.

*2.3.1. Effect of droplet's morphology on early stages of droplet growth*

Analysis of the early stages of droplet growth reveals that Cassie and partial Cassie droplets grow differently. Cassie droplet (**Figure 3e**) starts to grow on the top side of the cone (droplet is in focus) spreading subsequently between tops of two (2.2 s), three (2.6 s), four (4.2 s) and finally occupying six cones arranged hexagonally (5 s). Droplet is reflecting the light during the growth and cones are not visible through the droplet when focused on the top of the cones indicating that droplet is resting on the cones' tops. Growth of partial Cassie droplet (**Figure 3f**) starts between bottoms of the cones (the focal plane is on the tops of the cones and droplet is out of focus) spreading and growing between four neighbouring cones (0 - 2 s). Upon reaching the critical volume, the droplet wets hexagon with one cone inside (2.2 s). At this stage the droplet is 'flat' because all cones can be seen through the droplet without optical distortions. Then the volume of the droplet starts to increase, and the droplet becomes darker (3 s). After 4.6 s of growth droplet starts to act as magnifying lens and subsequently three (4.6 s), four (5.4 s), 7 (9.2 s) and more cones (13 s) can be easily seen. When droplet grow further (after 16.4 s), it becomes dark, and cones are no longer visible. For our system wetted area remains pinned when droplet starts to inflate upwards that aligns with three-step inflation model.[14] Optical observations are fully confirming ESEM results on droplet growth (**Figure S8e, f**).

*2.3.2. Effect of droplet morphology on individual droplet growth rate*

Since Cassie and partial Cassie droplets have different liquid–solid contact areas, their morphology strongly influences growth dynamics. Individual growth analysis revealed that partial Cassie droplets grow faster than Cassie droplets (**Figure S14**). This effect was most pronounced on sparse cones but was also present on dense cones due to the larger contact area of partial Cassie droplets. Overall, growth rate analysis across all samples (**Figure S15**) showed that dense cones exhibited nearly identical rates, while for sparse cones, the growth rate increased with cone spacing up to 3 μm.

The difference in growth rates on dense and sparse cones can be explained by the presence of noncondensable gases (NCGs), which create a water vapor depletion layer that hinders individual droplet growth. Due to NCGs, droplets located closer together on dense cones grow more slowly than isolated droplets on sparse cones[33]. Interestingly, for 4 μm spacing, where droplets predominantly adopted the Wenzel state, growth was slower than at 3 μm. We attribute this to the need for droplets in the Wenzel state to fill cone cavities before growing upward, whereas partial Cassie droplets only occupy a few cells before vertical growth begins.

*2.3.3. Droplet Morphology Distribution*

The proportion of different droplet types (**Figure 3g**) was defined by focal-plane scanning in an area of at least 0.62 mm$^2$ free from defects and contamination after 30 minutes of the condensation experiment. On the dense cones, the Cassie state dominates. As the spacing increases, the partial Cassie state becomes dominant and still allows droplet jumping. Finally, when the spacing equals the cone height, the Wenzel state becomes dominant and droplets are no longer able to jump. For the sample with 1.5 μm spacing, the partial Cassie state is dominant, unlike for 2 μm, where both partial Cassie and Cassie states are present. Otherwise, the trend of increased dominance of the partial Cassie state with increased spacing is valid for all other samples.

## 2.4. Design parameters of superhydrophobic microstructures for different applications

### *2.4.1. General considerations about surface design and coating properties*

Jumping condensation is sensitive to many parameters including: (i) surface topography design,[2,3,13,15,17] (ii) nucleation density,[13,29] (iii) environmental conditions,[13] and others. For sustained jumping condensation, droplets must remain highly mobile, in partial Cassie or full Cassie states, while avoiding transition to the Wenzel state, which prevents jumping and causes strong pinning as on a planar surface.[9] We created a design map and included data from our microconical system (**Figure 4a**) taking into account nucleation density in relation to spacing between surface feature's spacing $<L>/p$, where $<L>$ is mean separation distances between droplet centres defined by Poisson distribution $< L >= 1/2\sqrt{N_s}$ and energy criterion $E^*$ reflecting geometrical aspect of the system.[29] The region where Cassie and partial Cassie morphologies dominate lies within the parameter space defined by $E^* \in [0, 1]$ and $<L>/p \geq 10$.

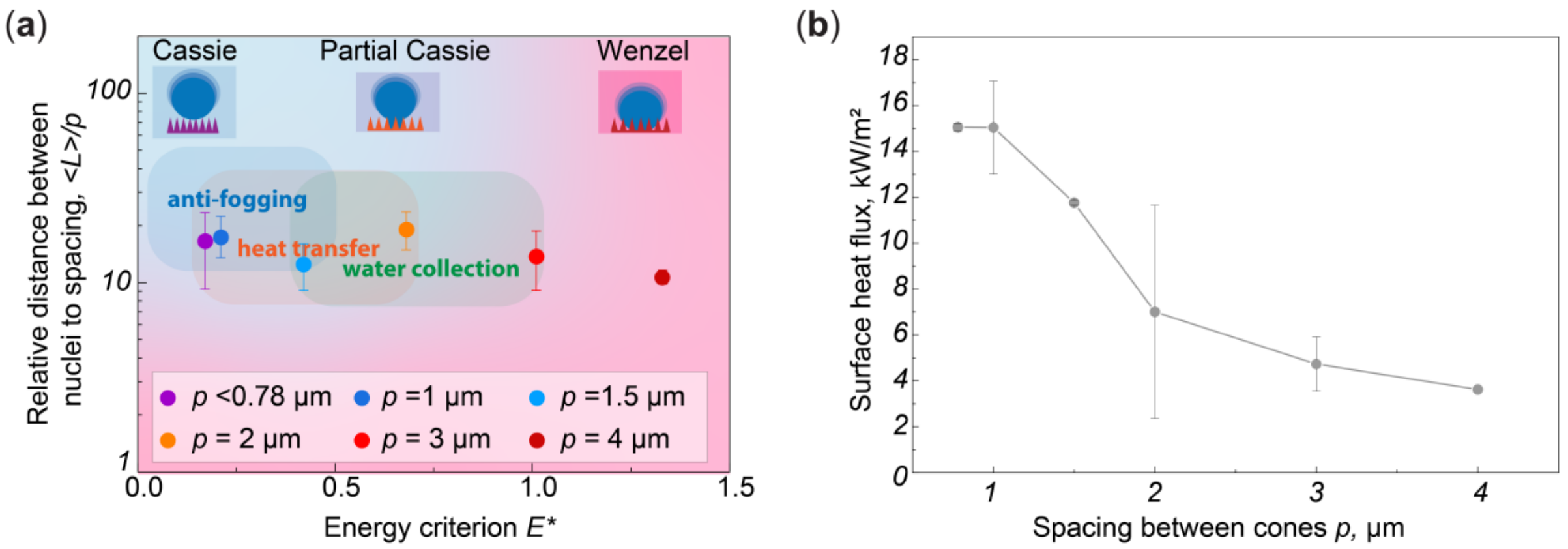

Figure 4. **Implications of droplet morphologies for different applications** (***a***) Surface design suggestions based on energy criterion $E^*$ and nucleation density in relation to spacing between surface feature's spacing $<L>/p$ resulting in different droplet morphology; (***b***) Surface heat flux in the vacuum chamber as function of spacing between cones $p$. Error bar is standard deviation calculated from at least 4 measurements ($n = 4\ldots20$).

The nucleation density, and thus $<L>/p$, governs both the Cassie–Wenzel transition[17,29] and the droplet jumping dynamics and efficiency. Across all our microconical samples, $<L>/p$ exceeded 10. However, we also observed transition to Wenzel state at $<L>/p \approx 1$ due to defects (**Figure S16a**), and these data were excluded from the analysis.

Among cones with 2 μm spacing, most samples exhibited a nucleation density $N_s \approx 160$ mm$^{-2}$ and $<L>/p \approx 20$ (**Figure S16b**), resulting in slow jumping dynamics ($N_j \approx 97$ mm$^{-2}$, $r_c \approx 15$ μm). In contrast, one sample (coated at the same time as the others) with $N_s \approx 720$ mm$^{-2}$ and $<L>/p \approx 10$ (**Figure S16c**), likely affected by submicron contamination, showed much more active jumping ($N_j = 616$ mm$^{-2}$, $r_c = 6$ μm). Surprisingly, in this latter sample most droplets were in the Cassie state, unlike on the other 2 μm samples (85 ± 9% vs. 40 ± 9%; **Figure S16c**). Considering this fact and the prevalence of partial Cassie droplets on the 1.5 μm sample (**Figure 3g**), we hypothesize that localized nanoscale defects (or contaminants) can shift the balance between Cassie and partial Cassie states. When defects are mainly located on the tops of the cones, nucleation occurs there, leading to the formation of Cassie droplets (**Figure 3e**). In contrast, defects at the base of the cones would promote nucleation at the base, resulting in partial Cassie droplets (**Figure 3f**). It should be noted that such an approach is effective when hydrophobic coating is uniform and localized defects are sparse enough ($<L>/p \geq 10$).

There is no consensus about effect of coating quality on nucleation density in conjunction with jumping condensation.[13,34] Contact angle hysteresis on planar surfaces is associated with chemical heterogeneities or/and roughness.[35] Low hysteresis is associated with uniform and smooth coating resulting in uniformly distributed and reduced nucleation density $N_s$ (on a planar surface).[24] Although low hysteresis on planar surfaces does not ensure low nucleation density on microstructures (**Figure S17**), we suggest to optimize the coating's deposition procedure for sustained jumping condensation.[36–38]

*2.4.2. Anti-fogging applications*

Anti-fogging refers to minimizing water condensation on surfaces cooled below the dew point in humid environments. Traditionally, a hydrophilic coating is used to reduce light scattering by forming a smooth, continuous water layer instead of discrete droplets.[39] However, this approach has drawbacks: (i) variable water film thickness lowers optical measurement accuracy, and (ii) coatings are easily contaminated. Alternatively, superhydrophobic surfaces exhibiting jumping condensation could repel condensing droplets limiting amount of water on the surface and droplet size.[2,3] Transparent anti-fogging surfaces can be achieved with nanoscale roughness that doesn't scatter visible light, while applications not requiring high level of transparency (e.g., infrared sensors, radio equipment, electronic enclosures) can use microstructured surfaces instead.

Anti-fogging based on jumping condensation strategies involve minimized nucleation density, surface coverage $\Phi$, and hence low amount of present condensed water on the surface $V$, minimal average droplet radius $<R>$, high jumping rate $J$ and number of jumps $N_j$. In terms of droplet morphology, Cassie droplets are preferable for anti-fogging applications since they do not renucleate after departure (**Figure 4b**) and have a slower droplet radius growth rate compared to partial Cassie droplets (**Figure S14**). Similar considerations about anti-fogging can be applied to anti-icing applications; however, we acknowledge that this application is more challenging. In the case of freezing condensing droplets, low surface coverage will be preferred, and dominance of Cassie droplets is preferable since they can be easily removed from the surface and will prevent the formation of ice bridging, known as a severe limitation of superhydrophobic surfaces for anti-icing.[40–43]

In terms of surface geometry, the choice of dense features with low $E^*$ is preferable. Nucleation density needs to be optimized in the range of $<L>/p \geq 10$ according to the application, since it will affect jumping dynamics. With very low nucleation density, a smaller number of droplets will be generated, but it will come with the cost of slower jumping dynamics and larger critical jumping droplet radius $r_c$ (see discussion about **Figure S13**). Local nanoscale defects on the top of the structures may alter the proportion of condensing droplets towards the Cassie state.

*2.4.3. Heat transfer*

Jumping condensation offers a promising route to enhance heat transfer by continuously refreshing the surface through droplet departure and renucleation, which lowers the time-averaged droplet size distribution on the condensing surface and thus reduces the overall thermal resistance from the vapor to the surface. Maximizing heat transfer performance requires sustained high nucleation site density, suppressing Wenzel-state transitions, and promoting both high jumping rate $J$ and number of jumps $N_j$.[44] Among various wetting states, the partial Cassie morphology is particularly favourable due to its higher droplet growth rate (**Figure S14**), robust renucleation capability (**Figure S12**), and superior single-droplet heat transfer performance compared to the Cassie state.[15] Maintaining a high nucleation density, characterized by $<L>/p \approx 10$, is critical for efficient heat transfer. While feature spacing can vary, ensuring $E^* < 1$ and increasing surface area via denser nanostructures promotes active jumping behavior through enhanced nucleation. Importantly, a homogeneous coating combined with uniformly distributed nanoscale defects at the cone bases favours stable partial Cassie droplets (**Figure 4a**), enabling sustained high heat transfer over multiple condensation cycles.

**Figure 4b** shows the predicted heat flux on different surfaces revealing a clear trend of decreasing heat flux with increasing spacing. The overall heat flux is calculated based on the proportions of different wetting states (**Figure 3g**), where the heat transfer for each wetting state is calculated using the specified model that combines the single-droplet heat transfer and the droplet size distribution, see details on **Supplementary Discussion 3** and **Figure S18**.

*2.4.4. Water collection*

Jumping condensation can also be beneficial for water collection.[11,43] By rational surface design, its performance can be optimized[43]. Since water collection happening in ambient atmosphere, presence of NCGs significantly affects condensation performance, see [43] and discussion above. The main goal is to maximize amount of water collected from the atmosphere during the phase change. Requirements for water collection are similar to heat transfer case, however critical jumping radius must not be too small to ensure that jumping droplet could be collected before evaporation. Therefore, features should have intermediate density, $E^{*}\approx 0.5$. Partial Cassie droplet morphology is also preferrable for collecting water for the same reasons as for the heat transfer. Maximum amount of condensed water (water trapped on the sample $V$ and removed by jumping $V_j$) was on the dense cones and 2 μm spacing (**Figure S7**). Critical jumping radii $r_c$ were 6.3±1 vs. 14.5±6.8 μm respectively (**Figure 2a)**, making 2 μm spacing more preferrable. Uniform coating ensures sustained water collection, preventing the transition to the Wenzel state that is associated with loss of performance.

## 3. Conclusions

This work demonstrates that microstructured surfaces can exhibit a spacing-dependent transition that fundamentally governs jumping condensation. By varying the pitch between microcones, we find that this single geometric parameter dictates the characteristic droplet-removal size and delineates the boundary between efficient jumping and suppressed departure, consistent with the onset of Wenzel wetting. These results further show that macroscopic wetting parameters alone are insufficient to predict condensation behavior, as microdroplets interact with the full three-dimensional geometry of the cones rather than only their tips.

The microconical architecture also enables direct optical identification of Cassie, partial Cassie, and Wenzel droplets via simple focal-plane scanning. Dense arrays favor Cassie droplets, while increasing spacing promotes partial Cassie states whose residual wet patches seed new nucleation. Local nanoscale defects, combined with a uniform low-surface-energy coating, additionally modulate these morphologies, revealing multiscale sensitivity in practical surfaces.

Altogether, the findings establish a quantitative framework linking microstructure spacing, droplet morphology, and nucleation density. The resulting geometry–wetting design map provides a rational pathway for engineering scalable, mechanically robust surfaces tailored for anti-fogging, heat-transfer enhancement, and water harvesting. This framework elevates microstructured surfaces from empirical designs to rationally engineered platforms for real-world condensation control.

## 4. Experimental Section

Details of all experiments are included in the Supporting Information.

**Acknowledgements**

The authors also acknowledge the provision of facilities and technical support by the Aalto University at the OtaNano Nanomicroscopy Center and OtaNano Micronova Nanofabrication Center, as well as Light Microscopy Unit, Institute of Biotechnology, supported by HiLIFE and Biocenter Finland. M.S.K. thanks I. Belevich for helping with ESEM imaging, D. Quéré for inspiring discussions, A. Opalev for automation of auxiliary tasks.

**Author Contributions**

M.S.K. designed the experiments and performed optical imaging of condensation including focal-plane scanning, SEM, ESEM. Optical imaging data were analyzed by M.S.K. with the help of M.L., T.K. developed and validated the magnification effect model, analyzed optical focal-plane scanning and performed confocal imaging. T.Kos. (Tomi Koskinen), R.R. and V.P.J. fabricated microstructured surfaces, which were coated by M.S.K. with aid of P.K. Wetting characterization was performed by M.S.K. and P.K. (CA of planar samples), S.L. (CA of microstructured samples), and M.V. (SDAM); J.L and N.M. contributed to heat transfer modelling. R.H.A.R., J.V.I.T., I.T. supervised experiments and sample preparation. M.S.K. prepared the original draft of the manuscript; all authors reviewed and approved the manuscript.

**Data Availability Statement**

The data that support the findings of this study are available from the corresponding authors upon reasonable request.